\documentclass[aps,prb,amsmath,amssymb,reprint,floatfix]{revtex4-1}

\usepackage{hyperref}
\usepackage{epsfig}
\usepackage{graphicx}
\usepackage{subfigure}
\usepackage{latexsym}
\usepackage{color}
\usepackage{fullpage}
\usepackage{dcolumn}
\usepackage{bm}
\usepackage{ulem}
\usepackage{units}
\usepackage{amsmath}

\begin{document}

\title{Effective interaction quenching in artificial kagomé spin chains}

\author{L. Salmon$^{1}$, V. Sch\'anilec,$^{1,2}$, J. Coraux$^{1}$, B. Canals$^{1}$, and N. Rougemaille$^{1}$}

\affiliation{$^{1}$ Univ. Grenoble Alpes, CNRS, Grenoble INP, Institut NEEL, 38000 Grenoble, France \\
$^{2}$ Central European Institute of Technology, CEITEC BUT, Brno University of Technology, Purky\v{n}ova 123, Brno 612 00, Czech Republic}

\date{\today}

\begin{abstract} 
Achieving thermal equilibrium in two-dimensional lattices of interacting nanomagnets has been a key issue on the route to study exotic phases in artificial frustrated magnets. 
We revisit this issue in artificial one-dimensional kagomé spin chains. 
Imaging arrested micro-states generated by a field demagnetization protocol and analyzing their pairwise spin correlations in real space, we unveil a non-equilibrated physics.  
Remarkably, this physics can be reformulated into an at-equilibrium one by rewriting the associated spin Hamiltonian in such a way that one of the coupling constants is quenched. 
We ascribe this effective behavior to a kinetic hinderance during the demagnetization protocol, which induces the formation of local flux closure spin configurations that sometimes compete with the magnetostatic interaction.
\end{abstract}

\maketitle

\section{Introduction}

Arrays of interacting magnetic nanostructures are powerful experimental platforms in which to probe and emulate a wide range of phenomena often associated with highly frustrated magnetism \cite{Nisoli2013, Rougemaille2019}. 
For example, spin ice physics \cite{Harris1997, Bramwell2020, Wang2006, Tanaka2006, Qi2008, Ladak2010, Mengotti2011, Rougemaille2011, Zhang2012, Chioar2014a}, and Coulomb phase properties more specifically \cite{Henley2010, Moessner2016, Perrin2016, Sendetskyi2016, Canals2016, Ostman2018b, Farhan2019, Schanilec2020, Brunn2021, Rougemaille2021, Schanilec2022, Yue2022, Hofhuis2022}, have become accessible to direct imaging using lithographically patterned arrays of nanomagnets.
More recently, other forms of thus-fabricated synthetic two-dimensional (2D) lattices have been designed to study a zoology of frustrated lattice spin models \cite{Heyderman2020} having no equivalent in bulk compounds \cite{Gilbert2014, Gilbert2016, Farhan2017, Louis2018, Marrows2018, Leo2018, Nisoli2021}.

\begin{figure*}
\centering
\includegraphics[width=112.38 mm]{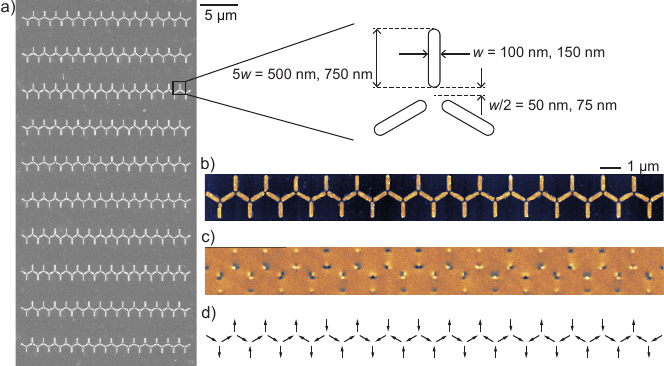}
\caption{(a) Scanning electron micrograph of a series of 10 kagomé spin chains and schematics illustrating the typical dimensions of the nanomagnets constituting the chains. (b) Atomic force microscopy image of an artificial kagomé spin chain. (c) Magnetic force microscopy image of the same chain after a field demagnetization protocol. The dark/bright contrasts give the magnetization direction within each nanomagnet. (d) From the magnetic image shown in (c) the spin micro-state of the chain is derived.}
\label{fig1} 
\end{figure*}

The properties of these artificially designed systems turn out to be usually well described by a physics at thermodynamic equilibrium.
In particular, all measurable quantities accessible through the imaging of spin micro-states can be rationalized with on-lattice spin models thermalized at a given temperature \cite{Nisoli2007, Nisoli2010, Chioar2014b, Perrin2019}. 
The experimental challenge then lies less in the ability to thermalize the system than in the capability to access different temperatures \cite{Schanilec2020, Schanilec2022}. 
This is particularly true when it comes to exploring physical phenomena that emerge at low temperatures or when approaching a phase transition \cite{Farhan2013, Anghinolfi2015, Sendetskyi2016, Sendetskyi2019, Hofhuis2022}. 

While the vast majority of studies to date have focused on 2D arrays of various geometries, sophisticated fabrication methods are being employed to build and image three-dimensional (3D) \cite{Ladak2019, Llandro2020, Ladak2021, Ladak2023} or quasi-3D \cite{Perrin2016, Farhan2019} artificial arrays. 
In contrast, there are few works only on one-dimensional (1D) systems, and these works have been devoted to the physics of Ising chains with ferromagnetic  \cite{Uppsala2018} or antiferromagnetic \cite{Dai2017} interactions. 
Interestingly, in contrast to their 2D counterparts, artificial spin chains sometimes exhibit richer phase diagram than predicted from their spin model. 
This is the case of the Ising spin chains with antiferromagnetic interactions \cite{Dai2017}, in which ordered patterns can be stabilized experimentally due to micromagnetic effects \cite{Rougemaille2013, Branford2013}, not accounted for by a point dipole model. 
These effects open up new prospects for the investigation of metastable configurations. 
They also illustrate the value of comparing the thermodynamics of spin models with their experimental emulation.

Pursuing this idea, we study in this work the magnetic correlations in kagomé spin chains submitted to a field demagnetization protocol, with the aim of understanding how their physics differs from that of their 2D parent lattices, for which an abundant literature exists\cite{Rougemaille2011, Tanaka2006, Qi2008, Ladak2010, Mengotti2011, Zhang2013, Montaigne2014, Farhan2017b, Dedalo2021, Kevin2020}. 
Following the methodology employed previously \cite{Dai2017}, we fabricated permalloy kagomé spin chains that were demagnetized using an external applied magnetic field. 
The resulting spin micro-states were then imaged using magnetic force microscopy. 
Analysis of the magnetic correlations reveals that these chains exhibit signatures of a non-equilibrated physics, not observed in 2D kagomé lattices. 
In particular, one magnetic correlator differs substantially and systematically from the predicted equilibrium value. 
Remarkably, this non-equilibrated physics can be rationalized using a modified spin Hamiltonian in which a specific coupling constant is quenched. 
The values of the correlations obtained with this modified Hamiltonian, thermalized at a well-defined temperature, nicely agree with the experimental measurements. 
In other words, while the field procedure applied to our kagomé spin chains does not allow us to obtain micro-states representative of configurations at thermodynamic equilibrium, an effective thermodynamics does account for the experimental data, provided that a certain coupling in the spin Hamiltonian is quenched. 
In this sense, artificial kagomé spin chains have a behavior upon field demagnetization that differs from that of their parent 2D lattices.
However, both systems share a common property: after being field demagnetized, they remain frozen at a relatively high temperature, of the order of a fraction of the coupling strength between nearest neighbors.
Thus, reducing the dimensionality does not appear as an efficient mean to reach low-energy configurations in demagnetized artificial systems.

\section{Experimental and theoretical details}

\subsection{Sample fabrication and demagnetization}

The kagomé spin chains are made of permalloy and the nanomagnets have $w \times 5w \times 30$ nm typical dimensions, with $w =$ 100 or 150 nm being the width of the nanomagnet [see Fig.~\ref{fig1}a)]. 
To ensure a strong magnetostatic coupling between the elements, the distance separating the extremity of a nanomagnet from the vertex center is fixed to $w/2$. 
The chain parameter is therefore $6w$. 
Each chain consists of $n=$29 (resp. 40) vertices, i.e., $N=59$ (resp. 81) nanomagnets when $w =$ 150 nm (resp. 100 nm). 
These chains were fabricated by electron lithography (lift-off) on a Si substrate. 

Following previous works \cite{Wang2007, Ke2008, Morgan2013}, the sample was rotated within an external magnetic field with an oscillating and slowly decaying amplitude, and direction within the sample plane. 
The applied magnetic field typically decays from $\pm$250 mT to 0 within 80 hours, while the sample rotation frequency is of the order of 10 Hz. 
The resulting magnetic configurations are imaged at room temperature using a magnetic force microscope.
Figures~\ref{fig1}b,c) show respectively a topographic and magnetic image of a given kagomé chain. 
In the magnetic image, dark and bright contrasts indicate the north and south poles. 
The absence of contrast between the two extremities of the nanomagnets indicates they are single domain, and their magnetic state can be approximated by an Ising variable [Fig.~\ref{fig1}d)].

\begin{figure}
\centering
\includegraphics[width=8cm]{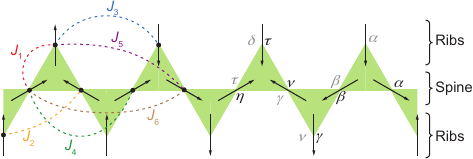}
\caption{Schematics of a kagomé spin chain showing the coupling strengths $J_k$ between the $k^\mathrm{th}$ neighbors and the notations used to identify pairwise spin correlations. The topology of the chain consists of a longitudinal spine bridging all vertices, and two transverse ribs composed of dangling spins.}
\label{fig2} 
\end{figure}

From the imaged magnetic configurations, spin-spin correlations can be measured. 
Pairwise spin correlations are defined as: $C_{ij}=1/N_{ij} \sum_{i \neq j} \sigma_i \sigma_j~\mathbf{e_i} \cdot \mathbf{e_j}$, where $\sigma_i$ is an Ising variable ($\sigma_i = \pm1$) residing on the site $i$ defined by the local unit vector $\mathbf{e_i}$, whereas $N_{ij}$ is the number of $ij$ pairs.
The correlations between the first six neighbors were computed from the measurements (the nomenclature for these correlations is illustrated in Fig.~\ref{fig2}: $C_{\alpha j}$ with $j=\beta,\gamma\cdots$ corresponding to $j=i+1,i+2\cdots$).
To improve the statistics, three series of 10 kagomé chains are considered in this work, two series ($\#1$ and $\#2$) for $w =$150 nm and one series ($\#3$) for $w =$100 nm, and each series has been demagnetized twice [one series is shown in Figure~\ref{fig1}a)]. 
Each correlation measurement is therefore the average of 20 chains consisting of 59 or 81 Ising variables.

  \begin{table}
 \caption{\label{table1}
Coupling strengths used in the dipolar spin Hamiltonian.
 }
 \begin{tabular}{|p{1cm}|p{1cm}|p{1cm}|p{1cm}|p{1cm}|p{1cm}|}
\hline
\hline
 $J_1$ & $J_2$ & $J_3$ & $J_4$ & $J_5$ & $J_6$ \\
\hline
1 & -0.137 & 0.045 & -0.036 & 0.014 & 0.037 \\
 \hline
 \hline
 \end{tabular}
 \end{table}

\subsection{Tensor matrix approach}

The correlations measured experimentally are then compared to those predicted by the thermodynamics of an on-lattice Ising spin Hamiltonian $\mathcal{H}_k$, which includes dipolar spin-spin interactions up to the $k^\mathrm{th}$ neighbors.
The Hamiltonian is thus of the form: $\mathcal{H}_k = \sum J_{ij} \sigma_i \sigma_j$, with $J_{ij}$ taking $k=1-6$ possible values (see Table.~\ref{table1}) according to the notation described in Fig.~\ref{fig2}.

Rather than using a Monte Carlo approach to probe the thermodynamics of our spin chains, we opted for an exact resolution using the transfer matrix method.
Here, we go beyond the textbook treatment known for 1D spin chains with magnetic interactions up to the third neighbor \cite{Zarubin2020}. 
More specifically, taking into account the coupling terms beyond the nearest-neighbors requires to define spin cluster matrices having $2N \times 2N$ dimensions, where $N$ is the number of spin comprised within the cluster. 
Resolution is obtained numerically and is not analytical, contrary to what was achieved for couplings up to the third neighbor \cite{Zarubin2020}.

\begin{figure*}
\centering
\includegraphics[width=16cm]{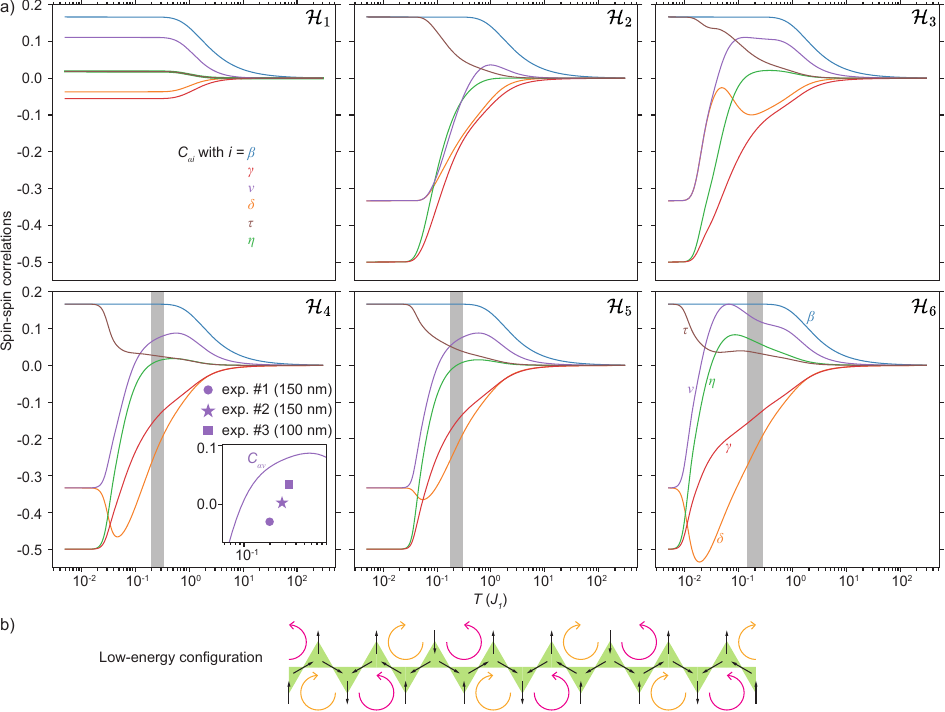}
\caption{(a) Temperature dependence of the magnetic correlations between the first six nearest neighbors as defined in Fig.~\ref{fig2}. Six scenarios are considered depending on the range of the dipolar interaction. From the upper-left panel to the lower-right panel, the range of the interaction extends from the first to the sixth neighbor, as indicated by the index $k$ in the spin Hamiltonian $\mathcal{H}_k$. The grey shaded area in the lower three panels indicates the temperature window compatible with the measured first three correlations ($C_{\alpha \beta}$, $C_{\alpha \gamma}$ and $C_{\alpha \delta}$). The inset in the case of $\mathcal{H}_4$ shows the discrepancy between the model and the experiment data for $C_{\alpha \nu}$ within the compatible temperature window. (b) Schematics of the configuration found at low temperature in all cases but $\mathcal{H}_1$. This configuration is the ground state found in the dipolar kagomé spin ice.}
\label{fig3} 
\end{figure*}

\begin{figure*}
\centering
\includegraphics[width=16cm]{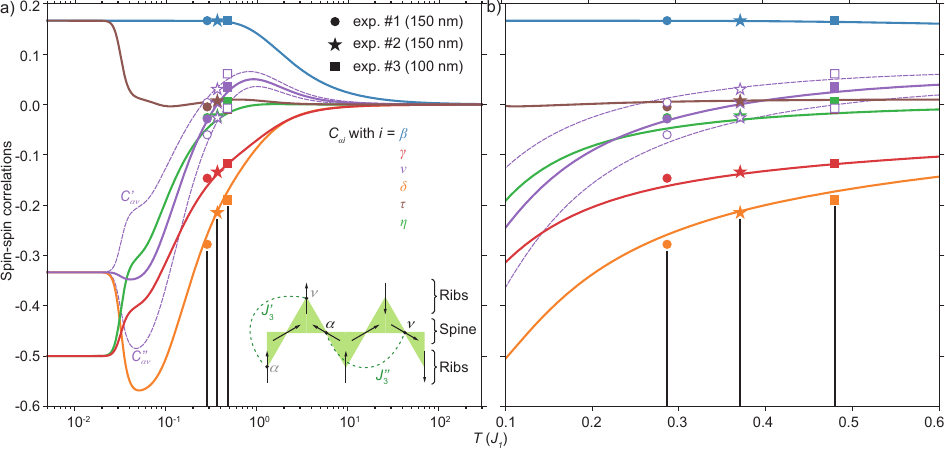}
\caption{(a) Temperature dependence of the magnetic correlations between the first six nearest neighbors as defined in Fig.~\ref{fig2} for the $\mathcal{H}_4$ spin Hamiltonian in which $J_3^{''}$ is set to 0. The predictions fit well fit the experimental findings marked by colored dots, stars and squares for samples $\#1$, $\#2$ and $\#3$, respectively. The difference between the $J_3^{'}$ and $J_3^{''}$ couplings is illustrated in the inset. The corresponding $C_{\alpha \nu}^{'}$ and $C_{\alpha \nu}^{''}$ correlators are plotted as purple dashed lines and open symbols. (b) Zoom-in of the full temperature window reported in (a).}
\label{fig4} 
\end{figure*}

 \begin{table}
 \caption{\label{table2}
Pairwise spin correlators measured for the first six neighbors in our three series of kagomé chains and from previous works on 2D kagomé lattices.
 }
\begin{ruledtabular}
 \begin{tabular}{|p{0.8cm}|p{1cm}|p{1cm}|p{1cm}|p{1cm}|p{1cm}|p{1cm}|}
    & $C_{\alpha \beta}$	& $C_{\alpha \gamma}$ & $C_{\alpha \delta}$ & $C_{\alpha \nu}$ & $C_{\alpha \tau}$ & $C_{\alpha \eta}$ \\
\hline
$\#$1    & 0.167 & -0.146 & -0.278 & -0.029 & -0.005 & -0.026 \\
\hline
$\#$2    & 0.167 & -0.134 & -0.215 & -0.004 & 0.007 & -0.022 \\
\hline
$\#3$     & 0.167 & -0.117 & -0.189 & 0.035 & -0.010 & 0.007 \\
\hline
\hline
2D \cite{Qi2008}    & 0.167 & -0.079 & -0.130 & 0.165 &  &  \\
\hline
2D \cite{Rougemaille2011}    & 0.164 & -0.056 & -0.063 & 0.151 & 0.013 & 0.056 \\
 \end{tabular}
 \end{ruledtabular}
 \end{table}

\section{Results}

The experimental measurements obtained for the three series of kagomé spin chains are reported in Table 1. 
Several observations can be made. 
First, the $C_{\alpha \beta}$ correlator is always equal to 1/6, demonstrating that the ice rule is strictly obeyed (no local 3-in or 3-out configuration is observed). 
Second, the $C_{\alpha \gamma}$ and $C_{\alpha \delta}$ correlators are significantly negative. 
If this is expected from magnetostatic considerations, we note that these two correlators are substantially larger in absolute value than what is found in 2D kagomé lattices \cite{Qi2008, Rougemaille2011}, typically by a factor 2 (see Table~\ref{table2}).
This result may suggest that a field demagnetization protocol brings more efficiently 1D kagomé chains into a low-temperature regime than 2D kagomé lattices.
Finally, the correlators involving the fourth, fifth and sixth neighbors ($C_{\alpha \nu}$, $C_{\alpha \tau}$ and $C_{\alpha \eta}$) are significantly lower in absolute value than the first three correlators, and their values are scattered around zero. 
Interestingly, these three last correlators, and $C_{\alpha \nu}$ in particular, markedly differ from the values reported in 2D networks (see Table~\ref{table2}), where they are systematically and significantly positive.
The question is then why the spin model describing chains and networks differs, and whether this is a dimensionality effect.

To address this question, we first consider the spin Hamiltonian $\mathcal{H}_k$ (we recall that $k$=1-6 is the range of the spin-spin interactions, up to the $k^\mathrm{th}$ neighbor) described in section II.B.
The temperature dependencies of the spin correlations deduced from the transfer matrix analysis are reported in Fig.~\ref{fig3}a) for the six  $\mathcal{H}_k$ Hamiltonians.
As expected, a nearest-neighbor description leaves the system macroscopically degenerate in its low-energy configuration, with spin correlations close to those predicted for 2D lattices \cite{Rougemaille2011, Wills2002}. 
In particular, once the ice rule is satisfied $(C_{\alpha \beta}=1/6)$, all correlators are temperature independent. 
Conversely, with an interaction cutoff radius beyond nearest neighbors ($\mathcal{H}_2$...$\mathcal{H}_6$), ordered patterns emerge at low temperatures. 
The presence of longer range couplings lifts the degeneracy observed when only $J_1$ is considered. 
Remarkably, the low-temperature magnetic pattern is identical for all $\mathcal{H}_k$ $(k>1)$, with the correlators reaching the same values whatever the Hamiltonian. 
We note that this low-energy configuration, shown in Fig.~\ref{fig3}b) is the one of the dipolar kagomé spin ice \cite{Moller2009, Chern2011}. 
 
The question that now arises is whether these models account for the experimental observations. 
To answer this question, it is instructive to compare the calculated temperature dependencies of the spin correlators to the values reported in Table 1. 
Interestingly, the measurements clearly show that $C_{\alpha \delta}$ is always smaller than $C_{\alpha \gamma}$, whatever the chain series we consider. 
These measurements are incompatible with the first three Hamiltonians $\mathcal{H}_1$, $\mathcal{H}_2$, and $\mathcal{H}_3$ [in these three cases, $C_{\alpha \gamma}$ is close, but smaller than $C_{\alpha \delta}$, see red and orange curves in the upper panels of Fig.~\ref{fig3}a)]. 
Conversely, the other three models ($\mathcal{H}_4$, $\mathcal{H}_5$, $\mathcal{H}_6$) predict $C_{\alpha\gamma}$ and $C_{\alpha\delta}$ values agreeing with the experimental ones within a certain temperature range, indicated by the shaded areas and corresponding curves [red and orange ones, see lower panels of Fig.~\ref{fig3}a)]. 
Nevertheless, within this temperature window, the agreement is poor with the other spin correlators. 
In particular, $C_{\alpha \nu}$ is not capture by the models [see inset in the bottom-left graph of Fig.~\ref{fig3}a)].
The results are similar for $C_{\alpha \eta}$ and $C_{\alpha \tau}$, whose experimental values are always lower than the predicted ones, and whose sign is sometimes opposite to the expected one. 
These observations suggest that a non-trivial mechanism is at work in our experiments.

At this point, it is worth mentioning that the $C_{\alpha \nu}$ correlator (like $C_{\alpha \beta}$) has a specific feature compared to the other correlators: it involves spin pairs that belong to the spine \textit{and} to the ribs of the chain (the other correlators involve spin pairs belonging either to the spine of the chain \textit{or} linking a spin of the spine and a dangling spin, but never both at the same time). 
The $C_{\alpha \nu}$ correlator can thus be calculated for each of the two sub-sets of spin pairs [see inset of Fig.~\ref{fig4}a)]. 
Experimental measurements of the corresponding correlators, $C'_{\alpha\nu}$ (involving dangling spins) and $C''_{\alpha\nu}$ (spins within the spine), reveals a surprising result: $C'_{\alpha\nu}$ is systematically larger than $C''_{\alpha\nu}$ (see Table~\ref{table3}). 
Recalling that these pairs of spins are coupled by a third-neighbor term $J_3$ (Fig.~\ref{fig2}), this experimental observation questions the relevance of a single $J_3$ value.
In the following, we define $J_3^{'}$ as the coupling involving dangling spins and $J_3^{''}$ involving a spin pair in the spine of the chain [see inset of Fig.~\ref{fig4}a)]. 
Since the $C_{\alpha \nu}$ correlator is smaller than the predicted value, it is reasonable to assume that $J_3$ is effectively smaller than what it should be.
Experimentally, $C_{\alpha \nu}^{'}>C_{\alpha \nu}^{''}$, suggesting that $J_{3}^{'}>J_3^{''}$.
We emphasize that, from a dipolar point of view, there is however no reason to envisage two distinct $J_3$ coupling terms, and such a distinction can only be effective.

 \begin{table}
 \caption{\label{table3}
Pairwise spin correlators measured for the two subfamilies of the third neighbors in our three series of kagomé chains. $J_3^{'}$ is the coupling term involving a spin pair within the spine of the chain and $J_3^{''}$ the one involving dangling bonds.
 }
 \begin{tabular}{|p{0.8cm}|p{1cm}|p{1cm}|}
\hline
\hline
    & $C_{\alpha \nu}^{'}$	 &  $C_{\alpha \nu}^{''}$ \\
\hline
$\#$1    & 0.004 & -0.061  \\
\hline
$\#$2    & 0.032 & -0.025  \\
\hline
$\#$3    & 0.062 & 0.008  \\
\hline
\hline
 \end{tabular}
 \end{table}

The correlators deduced from the models studied previously (Fig.~\ref{fig3}) were thus recalculated by imposing $J_{3}^{''}=0$, leaving $J_{3}^{'}=J_3=0.045$.
The choice to set $J_{3}^{''}=0$ is simply based on the fact that $J_3$ is already small (see Table.~\ref{table1}), a natural smaller value being 0.
Unsurprisingly, the modified $\mathcal{H}_1$, $\mathcal{H}_2$ and $\mathcal{H}_3$ models are still not compatible with the experimental data. 
However, good agreement is obtained as soon as the model includes interactions up to at least the fourth neighbor ($\mathcal{H}_4$, $\mathcal{H}_5$ and $\mathcal{H}_6$ to a lesser extent).
Figure.~\ref{fig4}a) reports the best case scenario, when considering the $\mathcal{H}_4$ model. 
Interestingly, beyond the six studied correlators, the distinction between $C_{\alpha \nu}^{'}$ and $C_{\alpha \nu}^{''}$ is also very well described [see purple dashed curves and open symbols in Figs.~\ref{fig4}a,b)]. 
This overall agreement between experiments and numerical predictions is actually very good for both $\mathcal{H}_4$ and $\mathcal{H}_5$, but less so when $J_6$ is taken into account. 
This is again consistent with the fact that the effective coupling involving a spin pair within the spine of the chain is reduced. 
The $J_6$ coupling being ferromagnetic and involving spin pairs within the spine (and only within the spine, see Fig.~\ref{fig2}), taking it into account degrades the agreement with the experiments. 
In other words, the best case scenario is found when the coupling terms $J_{3}^{''}$ and $J_6$, involving spin pairs belong to the spine of the chain, are quenched. 
We note that agreement with experiments can be further (slightly) improved, by setting the $J''_3$ coupling to an intermediate value, between 0 and $J_3$ = 0.045 (nominal dipolar value), of $J''_3\sim J_3/3$.

\begin{figure}[h!]
\centering
\includegraphics[width=8cm]{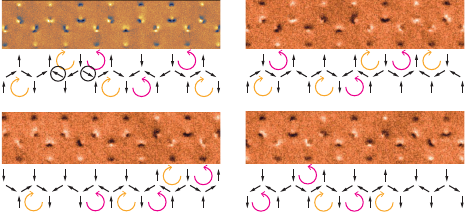}
\caption{Experimental magnetic images and their corresponding spin states for four different kagomé spin chains. Each chain is characterized by several local flux closure configurations represented by clockwise (yellow) and anticlockwise (purple) arrows. Successive yellow / purple loops impose an antiferromagnetic alignment of $\alpha \nu$ spin pairs belonging to the spine of the chain (two of such spins are marked by a black circle).}
\label{fig5} 
\end{figure}

\section{Interpretation}

The natural question that arises is why a modified Hamiltonian, in which $J_{3}^{''}$ is quenched, is able to account for the experimental measurements. 
It is striking that this is the case for three series of chains, found to be equilibrated at different effective temperatures [see Figs.~\ref{fig4}a,b)]. 
Although these effective temperatures lie in a narrow window (from 0.3 to 0.5 $J_1$ typically), some correlators ($C_{\alpha \delta}$ in particular) vary abruptly within this range, whereas others (like $C_{\alpha \tau}$) are essentially constant.
The Hamiltonian in which $J_{3}^{''}$ is quenched captures well all these temperature dependencies.

We stress again that the couplings are experimentally ruled by magnetostatics, and there is no reason why some couplings should be screened. 
For reasons likely originating from the demagnetization procedure, the kagomé chains are not found in configurations representative of thermodynamic equilibrium, but instead exhibit non-equilibrium behavior.
The quenching of a specific coupling can only be understood in an effective manner. 

Given the fact that the spin dynamics is governed by magnetization reversal processes, it may seem relevant to consider two coupling terms $J_3^{'}$ and $J_{3}^{''}$.
Indeed, a spin pair in the spine of the chain involves nanomagnets having a coordination number of 4, whereas it is 2 for spins involving dangling bonds.
Since magnetization reversal is triggered at the nanomagnet extremities, we might expect that dangling spins fluctuate more on average because of their free end than spins within the spine. 
We could then argue that dangling spins fluctuate sufficiently to reach an equilibrated state, whereas the dynamics of the spins belonging to the spine freezes quickly, being weakly correlated, especially when the associated coupling strength is small (like $J_3$).
More generally, we could argue that the physics of our spin chains is essentially governed by $J_{ij}$ couplings involving at least one dangling spin (with the exception of $J_{\alpha \beta}$, for which the interaction is very strong due to the small distance between the extremities of neighboring nanomagnets). 
Since only $J_{3}^{''}$ and $J_6$ couple spins all belonging to the spine of the chain, these two couplings could be effectively quenched because of the intrinsic field-induced magnetization dynamics within the chains.

We may also argue that the demagnetization procedure favors the formation of local flux-closure configurations.
Indeed, the spin micro-states we image clearly reveal local windings, sometimes over several successive half-hexagons.
This effect is illustrated in Fig.~\ref{fig5} for four chains in which local flux closure configurations are indicated by yellow and pink oriented loops.
Importantly, the formation of these local loops imposes a negative $C_{\alpha \nu}$ correlation along the spine (antiparallel spins), competing with the positive correlations expected from the magnetostatic interaction.
We note that the formation of such local loops is not incompatible with the argument above suggesting that dangling spins fluctuate more than spins within the spine of the chain.
In fact, these two effects could be intimately related, explaining why $J_{3}^{''}$ is effectively quenched.

\section{Conclusion}

In this work, we investigated magnetic correlations in field-demagnetized artificial kagomé spin chains.
The central result is the deviation of some of the experimental pairwise spin correlations from the values predicted at thermodynamic equilibrium, indicating that an out-of-equilibrium physics is a play in our experiment.
Like in antiferromagnetically-coupled Ising chains, the arrested micro-states we imaged are not predicted by the canonical spin Hamiltonian.
However, while in antiferromagnetic Ising chains an ordered pattern emerges, even in metastable configurations \cite{Dai2017}, the kagomé spin chains remain disordered.
In both kinds of chains, unpredicted configurations can be reached after a field protocol, opening new prospects to access metastable and/or unusual correlations in artificial spin systems.

Our results also suggest that the physics of small size 2D kagomé spin ice might be affected by the presence of dangling spins at the lattice edges.
In particular, magnetic correlations might be different in the bulk of the system and at the vicinity of the lattice edges.
To what extent such effects could impact the physics in 2D networks remains an open question.

\bigskip

\textit{Acknowledgments --}
The authors warmly thank Yann Perrin and Van-Dai Nguyen for providing technical help in the sample fabrication and characterization.


\bigskip



\begin{thebibliography}{}

\bibitem{Nisoli2013}
C. Nisoli, R. Moessner, and P. Schiffer,
Rev. Mod. Phys. \textbf{85}, 1473 (2013).

\bibitem{Rougemaille2019}
N. Rougemaille and B. Canals,
Eur. Phys. J. B \textbf{92}, 62 (2019).

\bibitem{Harris1997}
M. J. Harris, S. T. Bramwell, D. F. McMorrow, T. Zeiske, K. W. Godfrey, 
Phys. Rev. Lett. \textbf{79}, 2554 (1997).

\bibitem{Bramwell2020}
S. T. Bramwell and M. J. Harris,
J. Phys.: Condens. Matter \textbf{32}, 374010 (2020).

\bibitem{Wang2006}
R.F. Wang, C. Nisoli, R.S. Freitas, J. Li, W. McConville, B.J. Cooley, M.S. Lund, N. Samarth, C. Leighton, V.H. Crespi, P. Schiffer,
Nature \textbf{439}, 303 (2006).

\bibitem{Tanaka2006}
M. Tanaka, E. Saitoh, H. Miyajima, T. Yamaoka, and Y. Iye, 
Phys. Rev. B \textbf{73}, 052411 (2006).

\bibitem{Qi2008}
Y. Qi, T. Brintlinger, and J. Cumings,
Phys. Rev. B \textbf{77}, 094418 (2008).

\bibitem{Ladak2010}
S. Ladak, D. E. Read, G. K. Perkins, L. F. Cohen, and W. R. Branford,
Nature Phys. \textbf{6}, 359 (2010).

\bibitem{Mengotti2011}
E. Mengotti, L. J. Heyderman, A. Fraile Rodríguez, F. Nolting, R. V. Hügli, H.-B. Braun,
Nature Phys. \textbf{7}, 68 (2011).

\bibitem{Rougemaille2011}
N. Rougemaille, F. Montaigne, B. Canals, A. Duluard, D. Lacour, M. Hehn, R. Belkhou, O. Fruchart, S. El Moussaoui, A. Bendounan, and F. Maccherozzi, 
Phys. Rev. Lett. \textbf{106}, 057209 (2011).

\bibitem{Zhang2012}
S. Zhang, J. Li, I. Gilbert, J. Bartell, M. J. Erickson, Y. Pan, P. E. Lammert, C. Nisoli, K. K. Kohli, R. Misra, V. H. Crespi, N. Samarth, C. Leighton, and P. Schiffer, 
Phys. Rev. Lett. \textbf{109}, 087201 (2012).

\bibitem{Chioar2014a}
I. A. Chioar, N. Rougemaille, A. Grimm, O. Fruchart, E. Wagner, M. Hehn, D. Lacour, F. Montaigne, and B. Canals, 
Phys. Rev. B \textbf{90}, 064411 (2014).

\bibitem{Henley2010}
C.L. Henley, 
Ann. Rev. Condens. Matter Phys. \textbf{1}, 179 (2010).

\bibitem{Moessner2016}
J. Rehn and R. Moessner,
Phil. Trans. R. Soc. A \textbf{374}, 20160093 (2016).

\bibitem{Perrin2016}
Y. Perrin, B. Canals, and N. Rougemaille,
Nature \textbf{540}, 410 (2016).

\bibitem{Sendetskyi2016}
O. Sendetskyi, L. Anghinolfi, V. Scagnoli, G. M\"{o}ller, N. Leo, A. Alberca, J. Kohlbrecher, J. Luning, U. Staub, and L. J. Heyderman,
Phys. Rev. B \textbf{93}, 224413 (2016).

\bibitem{Canals2016}
B. Canals, I. A. Chioar, V. D. Nguyen, M. Hehn, D. Lacour, F. Montaigne, A. Locatelli, T. O. Mente\c{s}, B. Santos Burgos, and N. Rougemaille,
Nat. Commun. \textbf{7}, 11446 (2016).

\bibitem{Ostman2018b}
E. \"{O}stman, H. Stopfel, I.-A. Chioar, U. B. Arnalds, A. Stein, V. Kapaklis, and B. Hj\"{o}rvarsson, 
Nat. Phys. \textbf{14}, 375 (2018).

\bibitem{Farhan2019}
A. Farhan, M. Saccone, C. F. Petersen, S. Dhuey, R. V. Chopdekar, Y.-L. Huang, N. Kent, Z. Chen, M. J. Alava, T. Lippert, A. Scholl, and S. van Dijken, 
Sci. Adv. \textbf{5}, eaav6380 (2019).

\bibitem{Schanilec2020}
V. Sch\'anilec, B. Canals, V. Uhl\'i\v{r}, L. Flaj\v{s}man, J. Sad\'ilek, T. \v{S}ikola, and N. Rougemaille,
Phys. Rev. Lett. \textbf{125}, 057203 (2020).

\bibitem{Brunn2021}
O. Brunn, Y. Perrin, B. Canals, and N. Rougemaille,
Phys. Rev. B \textbf{103}, 094405 (2021).

\bibitem{Rougemaille2021}
N. Rougemaille and B. Canals,
Appl. Phys. Lett. \textbf{118}, 112403 (2021).

\bibitem{Schanilec2022}
V. Sch\'anilec, O. Brunn, M. Hor\'a\v{c}ek, S. Kr\'atk\'y, P. Meluz\'in, T. \v{S}ikola, B. Canals, and N. Rougemaille,
Phys. Rev. Lett. \textbf{129}, 027202 (2022).

\bibitem{Yue2022}
W.-C. Yue, Z. Yuan, Y.-Y. Lyu, S. Dong, J. Zhou, Z.-L. Xiao, L. He, X. Tu, Y. Dong, H. Wang, W. Xu, L. Kang, P. Wu, C. Nisoli, W.-K. Kwok, and Y.-L. Wang,
Phys. Rev. Lett. \textbf{129}, 057202 (2022).

\bibitem{Hofhuis2022}
K. Hofhuis, S. H. Skjærvø, S. Parchenko, H. Arava, Z. Luo, A. Kleibert, P. M. Derlet, and L. J. Heyderman,
Nat. Phys. \textbf{18}, 699 (2022).

\bibitem{Heyderman2020}
S. H. Skjærvø, C. H. Marrows, R. L. Stamps, and L. J. Heyderman,
Nat. Rev. Phys. \textbf{2}, 13 (2020).

\bibitem{Gilbert2014}
I. Gilbert, G.-W. Chern, S. Zhang, L. O’Brien, B. Fore, C. Nisoli, and P. Schiffer, 
Nat. Phys. \textbf{10}, 670 (2014).

\bibitem{Gilbert2016}
I. Gilbert, Y. Lao, I. Carrasquillo, L. O’Brien, J. D. Watts, M. Manno, C. Leighton, A. Scholl, C. Nisoli, and P. Schiffer,
Nat. Phys. \textbf{12}, 162 (2016).

\bibitem{Farhan2017}
A. Farhan, C. F. Petersen, S. Dhuey, L. Anghinolfi, Q. H. Qin, M. Saccone, S. Velten, C. Wuth, S. Gliga, P. Mellado, M. J. Alava, A. Scholl, and S. van Dijken,
Nat. Commun. \textbf{8}, 995 (2017).

\bibitem{Louis2018}
D. Louis, D. Lacour, M. Hehn, V. Lomakin, T. Hauet, and F. Montaigne,
Nat. Mater. \textbf{17}, 1076 (2018).

\bibitem{Marrows2018}
D. Shi, Z. Budrikis, A. Stein, S. A. Morley, P. D. Olmsted, G. Burnell, and C. H. Marrows,
Nat. Phys. \textbf{14}, 309 (2018).

\bibitem{Leo2018}
N. Leo, S. Holenstein, D. Schildknecht, O. Sendetskyi, H. Luetkens, P. M. Derlet, V. Scagnoli, D. Lançon, J. R. L. Mardegan, T. Prokscha, A. Suter, Z. Salman, S. Lee, and L. J. Heyderman,
Nat. Commun.  \textbf{9}, 2850 (2018).

\bibitem{Nisoli2021}
X. Zhang, A. Duzgun, Y. Lao, S. Subzwari, N. S. Bingham, J. Sklenar, H. Saglam, J. Ramberger, J. T. Batley, J. D Watts, D. Bromley, R. V. Chopdekar, L. O’Brien, C. Leighton, C. Nisoli, and P. Schiffer,
Nat. Commun. \textbf{12}, 6514 (2021).

\bibitem{Nisoli2007}
C. Nisoli, R. Wang, J. Li, W. McConville, P. Lammert, P. Schiffer, and V. Crespi, 
Phys. Rev. Lett. \textbf{98}, 217203 (2007).

\bibitem{Nisoli2010}
C. Nisoli, J. Li, X. Ke, D. Garand, P. Schiffer, and V. H. Crespi,
Phys. Rev. Lett. \textbf{105}, 047205 (2010).

\bibitem{Perrin2019}
Y. Perrin, B. Canals, and N. Rougemaille, 
Phys. Rev. B \textbf{99}, 224434 (2019).

\bibitem{Chioar2014b}
I. A. Chioar, B. Canals, D. Lacour, M. Hehn, B. Santos Burgos, T. O. Mentes, A. Locatelli, F. Montaigne, and N. Rougemaille, 
Phys. Rev. B \textbf{90}, 220407(R) (2014).

\bibitem{Farhan2013}
A. Farhan, P. M. Derlet, A. Kleibert, A. Balan, R. V. Chopdekar, M. Wyss, L. Anghinolfi, F. Nolting, L. J. Heyderman,
Nat. Phys. \textbf{9}, 375 (2013).

\bibitem{Anghinolfi2015}
L. Anghinolfi, H. Luetkens, J. Perron, M. G. Flokstra, O. Sendetskyi, A. Suter, T. Prokscha, P. M. Derlet, S. L. Lee, and L. J. Heyderman, 
Nat. Commun. \textbf{6}, 8278 (2015).

\bibitem{Sendetskyi2019}
O. Sendetskyi, V. Scagnoli, N. Leo, L. Anghinolfi, A. Alberca, J. Lüning, U. Staub, P. M. Derlet, and L. J. Heyderman,
Phys. Rev. B  \textbf{99}, 214430 (2019).

\bibitem{Ladak2019}
A. May, M. Hunt, A. van den Berg, A.Hejazi, and S. Ladak,
Commun. Phys. \textbf{2}, 13 (2019).

\bibitem{Llandro2020}
J. Llandro, D. M. Love, A. Kov\'acs, J. Caron, K. N. Vyas, A. K\'akay, R. Salikhov, K. Lenz, J. Fassbender, M. R. J. Scherer, C. Cimorra, U. Steiner, C. H. W. Barnes, R. E. Dunin-Borkowski, S. Fukami, and H. Ohno,
Nano Lett. \textbf{20}, 3642 (2020).

\bibitem{Ladak2021}
A. May, M. Saccone, A. van den Berg, J. Askey, M. Hunt, and S. Ladak,
Nat. Commun. \textbf{12}, 3217 (2021).

\bibitem{Ladak2023}
M. Saccone, A. Van den Berg, E. Harding, S. Singh, S. R. Giblin, F. Flicker, and S. Ladak,
Commun. Phys. \textbf{6}, 217 (2023).

\bibitem{Uppsala2018}
E. Östman, U. B. Arnalds, V. Kapaklis, A. Taroni, and B. Hjörvarsson,
J. Phys.: Condens. Matter \textbf{30} 365301 (2018).

\bibitem{Dai2017}
V.-D. Nguyen, Y. Perrin, S. Le Denmat, B. Canals, and N. Rougemaille,
Phys. Rev. B \textbf{96}, 014402 (2017).

\bibitem{Rougemaille2013}
N. Rougemaille, F. Montaigne, B. Canals, M. Hehn, H. Riahi, D. Lacour, J.-C. Toussaint, 
New J. Phys. \textbf{15}, 035026 (2013).

\bibitem{Branford2013}
K. Zeissler, S.K. Walton, S. Ladak, D.E. Read, T. Tyliszczak, L.F. Cohen, W.R. Branford, 
Sci. Rep. \textbf{3}, 1252 (2013).

\bibitem{Zhang2013}
S. Zhang, I. Gilbert, C. Nisoli, G.-W. Chern, M. J. Erickson, L. OBrien, C. Leighton, P. E. Lammert, V. H. Crespi, and P. Schiffer, 
Nature \textbf{500}, 553 (2013).

\bibitem{Montaigne2014}
F. Montaigne, D. Lacour, I.A. Chioar, N. Rougemaille, D. Louis, S. Mc Murtry, H. Riahi, B. Santos Burgos, T.O. Mentes, A. Locatelli, B. Canals, and M. Hehn, 
Sci. Rep. \textbf{4}, 5702 (2014).

\bibitem{Farhan2017b}
A. Farhan, P. M. Derlet, L. Anghinolfi, A. Kleibert, and L. J. Heyderman,
Phys. Rev. B \textbf{96}, 064409 (2017).

\bibitem{Dedalo2021}
D. Sanz-Hern\'{a}ndez, M. Massouras, N. Reyren, N. Rougemaille, V. Sch\'{a}nilec, K. Bouzehouane, M. Hehn, B. Canals, D. Querlioz, J. Grollier, F. Montaigne, and D. Lacour,
Adv. Mat. \textbf{33}, 2008135 (2021).

\bibitem{Kevin2020}
K. Hofhuis, A. Hrabec, H. Arava, N. Leo, Y.-L. Huang, R. V. Chopdekar, S. Parchenko, A. Kleibert, S. Koraltan, C. Abert, C. Vogler, D. Suess, P. M. Derlet, L. J. Heyderman,
Phys. Rev. B \textbf{102}, 180405(R) (2020).

\bibitem{Wang2007}
R. F. Wang, J. Li, W. McConville, C. Nisoli, X. Ke, J. W. Freeland, V. Rose, M. Grimsditch, P. Lammert, V. H. Crespi, and P. Schiffer,
J. Appl. Phys. \textbf{101}, 09J104 (2007).

\bibitem{Ke2008}
X. Ke, J. Li, C. Nisoli, Paul E. Lammert, W. McConville, R. F. Wang, V. H. Crespi, and P. Schiffer,
Phys. Rev. Lett. \textbf{101}, 037205 (2008).

\bibitem{Morgan2013}
J. P. Morgan, A. Bellew, A. Stein, S. Langridge, and C. H. Marrows,
Front. Physics \textbf{1}, 28 (2013).

\bibitem{Zarubin2020}
A. V. Zarubin, F. A. Kassan-Ogly, and A. I. Proshkin,
J. Magn. Magn. Mater. \textbf{514}, 167144 (2020).

\bibitem{Wills2002}
A. S. Wills, R. Ballou, and C. Lacroix, 
Phys. Rev. B \textbf{66}, 144407 (2002).

\bibitem{Moller2009}
G. Möller and R. Moessner,
Phys. Rev. B \textbf{80}, 140409 (2009).

\bibitem{Chern2011}
G.-W. Chern, P. Mellado, and O. Tchernyshyov, 
Phys. Rev. Lett. \textbf{106}, 207202 (2011).








\end{thebibliography}
\end{document}